# Astronomy should be in the clouds

*Arfon M. Smith <arfon@stsci.edu> STScI, Rob Pike <r@google.com> Google, William O'Mullane <womullan@lsst.org> LSST, Frossie Economou <frossie@lsst.org> LSST, Adam Bolton <bolton@noao.edu> NOAO, Ivelina Momcheva <imomcheva@stsci.edu> STScI, Amanda E Bauer, <abauer@lsst> AURA/LSST, Bruce Becker <brucellino@protonmail.ch> UEFA DevOps, Eric Bellm <ecbellm@uw.edu> University of Washington, Andrew Connolly <ajc@astro.washington.edu> University of Washington, Steven M. Crawford <scrawford@stsci.edu> STScI, Nimish Hathi <nhathi@stsci.edu> STScI, Peter Melchior <melchior@astro.princeton.edu> Princeton University, Joshua Peek <jegpeek@stsci.edu> STScI/JHU, Arif Solmaz <arifsolmaz@cag.edu.tr> Çağ University, Ross Thomson <drj@google.com> Google, Erik Tollerud <etollerud@stsci.edu> STScI, David W. Liska <dliska@stsci.edu> STScI*

Infrastructure Activity, Technological Development Activity, State of the Profession Consideration

**350 character summary:** Commodity cloud computing makes it possible for science projects to easily procure highly-reliable data management infrastructure on demand. In this paper we argue that astronomy should outsource our data management needs to the commercial cloud thereby allowing us to focus on our core competencies of data calibration and science exploitation.

## 1. Executive Summary

Commodity cloud computing, as provided by commercial vendors such as Amazon, Google, and Microsoft, has revolutionized computing in many sectors. With the advent of a new class of big data, public access astronomical facility such as LSST, DKIST, and WFIRST, there exists a real opportunity to combine these missions with cloud computing platforms and fundamentally change the way astronomical data is collected, processed, archived, and curated. Making these changes in a cross-mission, coordinated way can provide unprecedented economies of scale in personnel, data collection and management, archiving, algorithm and software development and, most importantly, science.

## 2. Recommendations

1. **Develop cloud-compatible business models:** Large facilities/missions should partner with commercial cloud vendors to develop cost-effective, science-enabling business models for astronomy that include placing high-value datasets in the cloud.
2. *Adopt a cloud-first approach for astronomy:* Major projects should demonstrate why they can't use the cloud, rather than assuming by default a non-cloud based custom infrastructure.
3. *Invest in Infrastructure as Code:* Facilities/missions/projects should invest in Infrastructure as Code[1] technologies to improve the portability and generalizability of their data management infrastructure.
4. **Develop common data management infrastructure components:** Large facilities/missions should coordinate to develop common Infrastructure as Code data management components that can be reused between missions.
5. **Provide sustained funding for training and cloud migrations:** Sustained funding should be made available to provide training and services for researchers and students wishing to use cloud computing and for projects wishing to migrate to the cloud.

## 3. Scientific potential of cloud computing today, and in the next decade

Two transformative developments will have a major effect on astronomical computation over the next decade. The first is the introduction of a new class of survey instruments, such as LSST, DKIST, and WFIRST, whose prime products are public datasets at a scale new to astronomy, commonly referred to as 'Big Data'. The second is the computing industry's shift towards a cloud-based commodity computing model (often referred to as Infrastructure as a Service, IaaS[2]), in which supercomputer centers and data centers of companies such as Amazon, Google, and Microsoft provide large-scale computational resources and services, as well as storage to host others' data at a competitive price. Together, these shifts represent a fundamental change in the way astronomical data can be collected, processed, archived, and curated, and if successfully harnessed, would offer enormous benefits for the accessibility and scientific utility of these datasets.

---

[1] https://infrastructure-as-code.com

## 3.1 The status quo

Historically, most astronomical data has been associated with the facility that collected it. This division of effort leads to multiple sets of similar (but crucially not the same) data management infrastructures, with each facility implementing its own data storage, networking services, computing/data processing components, and distribution infrastructure needed to support their projects and missions. In addition, these facilities typically fund an IT operations team responsible for keeping the system online. While efforts such as the Virtual Observatory (IVOA) have resulted in a modest number protocols and standards that, when implemented, allow for datasets to be queried in a uniform fashion between missions, 'under the hood', each center/facility essentially operates its own custom IT deployment which, while not astronomy-specific, nevertheless consume astronomy dollars.

When data management infrastructures represent a relatively modest fraction of the total mission budget, any duplication of effort and other inefficiencies are easy to overlook. At the petabyte scale, where data management budgets are much larger (~20% of the LSST construction budget is being spent on data management) the traditional model begins to look far from optimal, especially as there are a number of science-impeding and cost-increasing effects:

**Rigidity of infrastructure:** When infrastructure purchases happen on a ~5 year timeframe, any choices made persist for many years, stifling innovation and making experimentation and enhancements hard for data management teams and increasing the cost of change in terms of downtime.

**Evolving data access model:** The current service model for the vast majority of missions is to provide query and download services to the community for astronomical data. For an increasing number of science use cases, especially involving data science techniques such as machine learning, simultaneous access to very large volumes of this data is required. If utilizing novel methods for data analysis requires first downloading significant volumes of data, then this presents a major barrier to many individuals, organizations, and groups, particularly those at traditionally under-served organizations.

---

[2] https://csrc.nist.gov/glossary/term/Infrastructure-as-a-Service

**Infrastructure 'dead end' for missions at end of life:** Missions that have reached end of life often can't benefit from improvements to data management technologies as the budgets for these efforts have typically all but ceased. As the scientific utility of historical/archival datasets is well understood (White et al, 2009, Szalay 2019), this 'freezing in time' of data management infrastructure for archival missions limits any extended scientific return on our investment as a community.

## 3.2 Cloud Computing and Infrastructure as Code technologies

Cloud computing/IaaS combined with the Infrastructure as Code (IaC) pattern offers a solution to many of the limitations faced with traditional models of IT infrastructure. IaC allows organizations describe arbitrarily complex IT infrastructure in the form of configuration files and machine-readable scripts that can be used to provision infrastructure on-demand in the cloud on a 'pay as you go' basis. Because of the 'on demand' nature of these infrastructures, they typically can scale up (and down) as required based on real time demand. Organizations at all scales and in all sectors[3] including notable names such as Netflix (for scalability)[4], NASA Earth Science (for increased innovation, cost saving, and flexibility)[5], and the CIA (for improved information security)[6] are adopting cloud computing and IaC as their preferred approach for building IT infrastructure.

## 3.3 How Cloud Computing and IaC can benefit astronomy

While a single project or center can benefit individually from these technologies, for example, the PHAT survey collaboration made extensive use of Amazon Web Services for their data processing needs (Williams et al. 2018), adopting cloud computing and IaC in a coordinated way between many projects can facilitate reuse and deliver improvements to the ways in which we work as a community that ultimately will provide a greater scientific return on our investments.

An example of this approach bearing fruit in the sciences are the cloud-native services for genomics provided by Amazon and Google[7]. These *Science as a Service* offerings have helped the Broad Institute routinely ingest more than 24TB of data per day, store more than 36 PB of data in the cloud, and reduce the cost of a full genome analysis from $45 to $5 (Sheffi, 2018). With a coordinated approach in astronomy, similar generalizable services could be developed.

---

[3] https://cloud.google.com/customers, https://aws.amazon.com/solutions/case-studies, https://azure.microsoft.com/en-us/case-studies/
[4] https://aws.amazon.com/solutions/case-studies/netflix/
[5] https://earthdata.nasa.gov/cmr-and-esdc-in-cloud
[6] https://fcw.com/articles/2017/06/14/cia-cloud-aws.aspx
[7] https://cloud.google.com/genomics, https://aws.amazon.com/health/genomics

In the IaC model, because any infrastructure deployment is described in a collection of scripts, a set of technologies developed by one center, project, or facility, can be easily replicated by others. A compatible set of cross-platform / cross-provider (Amazon, Google, Microsoft) IaC deployments that enable advanced data processing capabilities for *all* observatory facilities, not just the few with extended resources, is well within reach when adopting this approach.

In the IaC and cloud computing paradigm, IT infrastructure is thought of as a *commodity resource* where key items such as data storage, computation, and networking are provided by a third party, but so are power, cooling, security, upgrades, and infrastructure IT staff for the data center. Over the past decade, commercial cloud vendors have become highly specialized in providing these services at enormous scale (Johnson, 2017), and extremely high levels of reliability.

**This ability to offer highly-reliable data management infrastructure on demand means the astrophysics community can focus on its core competencies such as developing novel algorithms for extracting science from mission data rather than building and maintaining custom data management infrastructure which is often inferior to, and more expensive than what is made available by Amazon, Google, and Microsoft.**

Below we discuss some key aspects of scientific data management and how we anticipate they would be affected by the adoption of cloud computing/IaaS and infrastructure as code technologies:

3.3.1 Data collection and management

As noted earlier, currently most astronomical data is associated with the facility that collected it. In a future more data-intensive regime, where a cloud-based model had been adopted for data management and processing, facilities would purchase (or possibly have donated - see the later section on challenges) resources to host their infrastructure in commodity cloud platforms. This would have a compelling number of positive consequences:

**More democratic, more efficient funding of computing resources for scientific analyses:** Commercial cloud computing makes it possible to rent a supercomputer by the hour (Descartes Labs, 2019) with individual researchers only paying for what they use. In these environments, access to the latest hardware (CPUs and GPUs) is available to all, rather than being available to a limited number of PIs and groups (Berman, 2019).

This 'levelling of the infrastructure playing field' becomes especially important for large survey instruments if the data are made available in a generally accessible facility such as the cloud. Astronomers who want to use the data can spend $10, $100, or $1000, as appropriate, to do their calculations and this model is vastly cheaper than building a data center in every school that wants to 'use' the survey instrument.

**Allow observatories/missions/facilities to focus on the unique aspects of their work:** Many data management activities generalize well between astronomy projects, and often scientific and industry workflows in general. For example, while data from different missions may have different characteristics, nearly every astronomical data archive offers a way to search for a list of sources in a particular region of the sky. Indeed with the advent of standardized protocols, these services are often invoked in the same way yet often are implemented with divergent, incompatible infrastructure 'back ends'. While specific algorithmic manipulations (e.g. calibration pipelines) are required for different data, the management of data processing pipelines and workflows is often remarkably generic yet almost always re-invented. Targeting generic commodity computing platforms for deployment of such services will naturally drive the field to share solutions for common issues (e.g. for storing and retrieving data during processing) while individual missions can focus on instrument- and calibration-specific development needs. Note that even in the absence of commercial cloud hosting, working with such prevalent industry paradigms promotes the use of standard design patterns that share many of the same benefits (Economou et al., 2014)

**Infrastructure reuse for smaller facilities & missions:** A cloud/IaC approach makes it significantly easier for smaller facilities and missions to reuse the data management infrastructure developed by larger entities, thus bringing the latest innovations in technologies, tools, and services to all projects.

**Efficiency of working 'server-side':** An IaaS/cloud model where individuals are able to run their own computation local to cloud-hosted missions datasets opens up the possibility of astronomers leveraging server-side analysis techniques. With the advent of computationally expensive statistical methods such as Bayesian inference, and applied machine learning techniques, this may be the only viable approach for many astronomers to perform their analyses on peta-scale datasets.

**High throughput computing:** By its nature, research has requirements that are intermittent: a particular analysis may have huge computation needs that, once complete, may leave machines idle. When suitably engineered in the cloud, one hour on 10,000 compute nodes is no more expensive than 10,000 hours on one. Building for this paradigm enables very high throughput analysis in IaaS/cloud environments by leveraging the massive scale offered by commercial cloud platforms[8].

**Lower barrier to entry and increased innovation:** When data management infrastructure is described in code, rapid prototyping of potential enhancements of existing infrastructure are significantly easier to develop as changes can be made in minutes and deployed and evaluated at very low marginal cost.

### 3.3.3 Astronomical archives

Cloud computing infrastructures amortize the tasks of maintaining massive data sets and of running the data centers, thereby simplifying the process of constructing and maintaining archives. While the cost of hosting and running a cloud-based archive needs to be fully investigated (see more on business models later), we believe that the cost of outsourcing will be offset by the savings from reduced personnel costs; shared computing, storage, and networking capacity; larger purchasing power; and other economies of scale.

**Computing and storage are becoming commodities. Astronomical facilities should be moving away from custom infrastructure deployments.**

### 3.3.4 Personnel

Analogous to the sequence of design, engineering, construction, commissioning, and operation of an astronomical facility, the large data sets that this new generation of instruments will have go through similar distinct phases. With multiple instruments at different stages of maturity, if projects follow similar, IaaS/cloud-based approaches to data management, personnel could more easily move between projects as their skills align with each project's current needs. There is a global shortage of experienced software developers of all skill sets, and this strategy would avoid the wasted time, expense, and competition of having each project recruit and train staff members from the same pool.

### 3.3.5 Education

Creating a community that can effectively use and extend a cloud-based infrastructure requires awareness and education. This necessity is in part responsible for the slow adoption of cloud computing within scientific communities (Bottum et al, 2017). Transitioning the community to work with the cloud as an integral part of the computing infrastructure will require sustained support from funding bodies. Training of different communities with the expertise to know what tools are available, and how to work with them becomes critical, and it is worth noting that such expertise is transferable across many types of data science

---

[8] https://mast-labs.stsci.io/2018/06/exploring-aws-lambda-with-hst-public-data

The current lack of such an educational path (formal or informal) creates a disconnect between the skills that are needed for an era rich in data and those that academia teaches to incoming graduate students and early-career postdocs. These challenges are discussed further in the Astro2020 white paper *"The Growing Importance of a Tech-Savvy Astronomy and Astrophysics Workforce"* by Norman et al. but in short, developing these tools and skills within the graduate student population and providing career paths for people with the necessary skill sets will be critical for maximizing the scientific return on big data missions, and is a role that could be fulfilled by a national lab or center.

### 3.3.6 Software development

While individual missions require new software to develop catalogs and other secondary or derived data products, much of the higher-order processing has a high degree of commonality across astronomical instruments. Even across disciplines as diverse as ground-based astronomy, space-based astronomy, solar physics, and radio astronomy there is the opportunity for significant sharing of software for construction, management, cataloging, and maintenance of data sets and catalogs. When working in the IaC paradigm, where infrastructure components are themselves described as code, the opportunity for reuse and cumulative ongoing development increases substantially.

## 3.4 Science-enabling opportunities of adopting the cloud

Key science of the 2020s is enabled by astronomy adopting the commercial cloud and infrastructure as code technologies:

**Multi-Messenger Astrophysics** relies upon near-real-time, on-demand scalable computing. In their Astro2020 decadal paper *Cyberinfrastructure Requirements to Enhance Multi-messenger Astrophysics*, Chang et al (2019) identify the need to access, analyze, and interpret large volumes of archival data from a broad collection of facilities and experiments. The reusable, reliable, and scalable nature of cloud plus IaC approaches described here make delivering on this need a realistic prospect in a cost-efficient, science-enabling way.

**Joint processing and repeated analyses across missions:** Many of the scientific merits of jointly processing LSST, WFIRST, and Euclid are described in Chary et al. 2019, Eifler et al. 2019, and Rhodes et al. 2019. Both LSST and WFIRST are already leveraging IaC technologies and planning to deploy parts of their data management infrastructure in the commercial cloud, which will make jointly-analysing these data a more tractable problem. In addition, a well described and implemented analysis using IaC will facilitate replication of analyses and results across efforts and missions.

**Stellar astrophysics in the era of large surveys** relies upon combining spectroscopy, photometry, astrometry, and variability across multiple ground-based and space-based data sets covering tens of millions to billions of stars and has a wide range of science applications (Huber et al. 2019, Sanderson et al. 2019, Williams et al. 2019, Rix et al. 2019). Staging datasets from these ground and space-based missions in the cloud would significantly improve the community's ability to jointly analyze these data products.

**Community science exploitation of big data missions, archival research & server-side analytics:** Big data survey missions such as LSST and WFIRST already assume a 'joint-responsibility data processing model' with the community data responsible for producing higher-level data products (L3 for LSST, L3/4/5 for WFIRST). In addition, archival data can be re-analyzed jointly with these upcoming missions. LSST is actively investigating deploying their LSST Science Platform environment to the cloud, citing scalability as a key factor (O'Mullane & Swinbank, 2018). Although earlier in development, WFIRST is likely to need similar capabilities for Science Investigation Teams (SITs) and serving the wider community needs and commercial cloud computing will likely play a major role here too.

**Maximizing the scientific potential of data science, machine learning, and AI:** Training of machine learning/AI methods such as deep learning algorithms often need simultaneous access to very large volumes of data and significant compute resources. Cloud-enabled science, where datasets are provided in an environment close to the essentially unlimited computational resources of the commercial cloud vendors would dramatically improve the ability of the astronomy community to leverage these techniques in their analyses.

**Facilitation of ad-hoc investigations by PIs:** Ready access to pre-configured IaaS data management infrastructure allows astronomers to immediately execute *ad hoc* data processing flows instead of waiting for a procurement cycle to acquire and provision a server through their home IT department. PIs who want to use the data can spend only what is needed to carry out their analyses. This approach is much cheaper than having a data center on every campus and enables even underfunded institutions to do new science.

**Reproducibility/replication:** In the cloud/IaC paradigm, the whole data management infrastructure behind any experiment or result (calibration pipelines/codes, machine environments, database schemas), is captured in a machine-readable form, thereby enabling reuse by others and more straightforward replication of work.

**More routine technology transfer from industry:** The commercial cloud combined with IaC is already the preferred approach for building advanced data science and data management infrastructures in industry. By using a similar toolchain to industry, astronomy can more easily adopt new data analysis techniques from industry and other technology transfer becomes much easier.

## 4. Potential challenges with this approach

While we believe the large-scale adoption of commercial cloud and IaC technologies in astronomy will allow us to go further and faster as a community in the next decade, but this change is not without challenges:

**Suitability of astronomical workloads to cloud environments:** Unless data management infrastructures are designed to be 'cloud native', moving from on-premise to the cloud can be expensive. Simple 'lift and shift' strategies, whereby the same infrastructure is (re)deployed to the cloud are rarely cheaper to operate and do not take advantage of the benefits of these platforms. Many astronomical data management workloads however are already a very good fit for the cloud: In the NSF-sponsored report, *'The Future of Cloud for Academic Research Computing'* Bottum et al (2017), 'pleasingly parallel' tasks such as data processing were identified as tasks/workloads that are already a very good fit for the commercial cloud and are often cheaper to run in such environments.

Traditional HPC workloads, such as generating large simulations, are very important to some areas of astronomy and astrophysics. Today, traditional HPC facilities are generally considered the most cost effective venue for HPC workloads. Given the rapid pace of innovation with commercial cloud vendors however, over the next 3-5 years, it's possible that the commercial cloud becomes a compelling choice for these workloads too (Bottum et al, 2017, Descartes Labs 2019).

**Storage costs and egress fees:** It is generally accepted that compute resources, especially for 'pleasingly parallel' computing workloads, can be procured in the cloud as cheaply as anywhere else. Where the commercial cloud can be cost-prohibitive, especially for projects with lots of data, is data storage and distribution (egress). In recent years, commercial cloud vendors have made a number changes that favor large science projects including 1) lowering of storage costs and offering different storage tiers, 2) public dataset programs for high-value scientific data[9] where significant quantities of data are stored at zero cost to the project (see HST[10] and TESS[11] as examples), 3) significant discounts for data downloads such as Google's egress waiver for Internet2 traffic to reduce costs[12], and 4) reduced pricing for archival storage. When properly architected, moving data between cloud storage and compute is extremely fast and very low cost. With many of the science drivers for big-data missions favoring a server-side analytics/code-to-data computing architecture, we believe cost efficient data management architectures can be developed.

---

[9] https://registry.opendata.aws, https://cloud.google.com/public-datasets
[10] https://registry.opendata.aws/hst/
[11] https://registry.opendata.aws/tess/
[12] https://cloud.google.com/billing/docs/how-to/egress-waiver

**Vendor lock in and reliance on commercial entities for data management:** One common concern raised about cloud computing is the so called 'vendor lock in' problem where, because of particular technology choices, organizations find themselves locked in/restricted to the technologies of a single vendor. A coherent approach to cloud computing across astronomy would enable the development of methodologies and the use of technologies (e.g. Kubernetes) that can be deployed across different cloud providers including academic clouds. The cost of vendor lock in for a particular technology (i.e. the cost to transition to a new cloud providers infrastructure) can then be assessed against the capabilities of the technology prior to its adoption by the astronomical community.

## 5. Conclusions

Whether it's the new class of astronomical facility such as LSST, DKIST, and WFIRST and their peta-scale archives, analyzing large volumes of data in real-time for multi-messenger astronomy, or applying novel computational data science techniques to archival holdings, the science of the next decade is going to either rely upon or be limited by data management and computing infrastructure. Peta-scale cloud computing infrastructure is not a core competency of academia and it shouldn't be: Commodity cloud computing infrastructure will allow us to tackle bigger problems, and scale far beyond what is possible with infrastructure developed by any single research discipline. Astronomy must learn to leverage these commodity technologies thereby allowing us to focus on our core competencies of data analysis, calibration, and science exploitation of these datasets.

With a coordinated approach, cloud computing platforms can usher in a new era in the way astronomical data is collected, processed, archived, and curated for major flagship projects, but also smaller missions too, and even individual PIs. Embracing this approach as a field will provide new opportunities for data exploration, allow for rapid and continual technology transfer from industry, and provide economies of scale in personnel, data collection and management, archiving, algorithm and software development and, most importantly, science.